\documentclass[letterpaper]{article} % DO NOT CHANGE THIS
\usepackage{aaai2026}  % DO NOT CHANGE THIS
\usepackage{times}  % DO NOT CHANGE THIS
\usepackage{helvet}  % DO NOT CHANGE THIS
\usepackage{courier}  % DO NOT CHANGE THIS
\usepackage[hyphens]{url}  % DO NOT CHANGE THIS
\usepackage{graphicx} % DO NOT CHANGE THIS
\usepackage{natbib}  % DO NOT CHANGE THIS AND DO NOT ADD ANY OPTIONS TO IT
\usepackage{caption} % DO NOT CHANGE THIS AND DO NOT ADD ANY OPTIONS TO IT
\usepackage{algorithm}
\usepackage{algorithmic}

\usepackage{times}
\usepackage{helvet}
\usepackage{courier}
\usepackage{xcolor}

\usepackage{amsmath}
\usepackage{url}            % simple URL typesetting
\usepackage{booktabs}       % professional-quality tables
\usepackage{amsfonts}       % blackboard math symbols
\usepackage{nicefrac}       % compact symbols for 1/2, etc.
\usepackage{microtype}      % microtypography
\usepackage{lipsum}
\usepackage{graphicx}
\graphicspath{ {./images/} }
\usepackage{amsmath}
\usepackage{tabularx}
\usepackage{multirow}
\usepackage{textcomp}
\usepackage{placeins}
\usepackage{newfloat}
\usepackage{listings}
\DeclareCaptionStyle{ruled}{labelfont=normalfont,labelsep=colon,strut=off} % DO NOT CHANGE THIS
\floatstyle{ruled}
\newfloat{listing}{tb}{lst}{}
\floatname{listing}{Listing}

\nocopyright

\title{Towards Hallucination-Free Music: A Reinforcement Learning Preference Optimization Framework for Reliable Song Generation}
\author{
 Huaicheng Zhang\textsuperscript{\rm 1}\thanks{Work performed during an internship at Tencent AI Lab.},
 Wei Tan\textsuperscript{\rm 2}\thanks{Correspinding Authors},
 Guangzheng Li\textsuperscript{\rm 2},
 Yixuan Zhang\textsuperscript{\rm 2},
 Hangting Chen\textsuperscript{\rm 2},
 Shun Lei\textsuperscript{\rm 3},
 Chenyu Yang\textsuperscript{\rm 4},
 Zhiyong Wu\textsuperscript{\rm 3},
 Shuai Wang\textsuperscript{\rm 5},
 Qijun Huang\textsuperscript{\rm 1$\dagger$},
 Dong Yu\textsuperscript{\rm 2$\dagger$}
}
\affiliations{
    \textsuperscript{\rm 1}Wuhan University, Wuhan, China\\
    \textsuperscript{\rm 2}Tencent AI Lab\\
    \textsuperscript{\rm 3}Shenzhen International Graduate School, Tsinghua University, Shenzhen, China\\
    \textsuperscript{\rm 4}The Chinese University of Hong Kong, Shenzhen (CUHK-Shenzhen), Shenzhen, China\\
    \textsuperscript{\rm 5}School of Intelligence Science and Technology, Nanjing University, Suzhou, China\\
    zhuaicheng@whu.edu.cn, waytan@tencent.com, huangqj@whu.edu.cn
}

\usepackage{bibentry}
\begin{document}

\maketitle

\begin{abstract}
Recent advances in audio-based generative language models have accelerated AI-driven lyric-to-song generation. However, these models frequently suffer from content hallucination, producing outputs misaligned with the input lyrics and undermining musical coherence. Current supervised fine-tuning (SFT) approaches, limited by passive label-fitting, exhibit constrained self-improvement and poor hallucination mitigation. To address this core challenge, we propose a novel reinforcement learning (RL) framework leveraging preference optimization for hallucination control. Our key contributions include: (1) Developing a robust hallucination preference dataset constructed via phoneme error rate (PER) computation and rule-based filtering to capture alignment with human expectations; (2) Implementing and evaluating three distinct preference optimization strategies within the RL framework: Direct Preference Optimization (DPO), Proximal Policy Optimization (PPO), and Group Relative Policy Optimization (GRPO). DPO operates off-policy to enhance positive token likelihood, achieving a significant 7.4\% PER reduction. PPO and GRPO employ an on-policy approach, training a PER-based reward model to iteratively optimize sequences via reward maximization and KL-regularization, yielding PER reductions of 4.9\% and 4.7\%, respectively. Comprehensive objective and subjective evaluations confirm that our methods effectively suppress hallucinations while preserving musical quality. Crucially, this work presents a systematic, RL-based solution to hallucination control in lyric-to-song generation. The framework's transferability also unlocks potential for music style adherence and musicality enhancement, opening new avenues for future generative song research.
\end{abstract}

\begin{links}
\link{Demo}{https://hallucination-free-music-demo.github.io/}
\end{links}


\section{Introduction}
Music has long been a fundamental aspect of human culture, serving as a universal language that transcends geographic and linguistic boundaries. With the rapid advancement of artificial intelligence generated content, the creation and composition of music have entered a new era, where algorithms can generate pure music, vocals, or songs. This technological evolution not only challenges traditional notions of creativity but also opens unprecedented opportunities for musicians, producers, and listeners. \par

Recently, natural language processing (NLP) has experienced a paradigm-shifting boom in both research and practical applications \cite{achiam2023gpt, touvron2023llama, sun2024hunyuan}. The demonstrated success of large language models (LLMs) in NLP has catalyzed their adoption in song generation \cite{yuan2025yue, bai2024seed, lei2025levo, gong2025ace} and text-to-speech (TTS) synthesis \cite{wang2023neural, du2024cosyvoice, ju2024naturalspeech}, driving a new phase of rapid technological advancement in these fields. Although LLM-based song generation achieves remarkable outcomes, such as style-consistent melody composition and context-aware lyric generation, it inherently suffers from hallucination phenomena. Hallucinations are considered incorrect or misleading generated contexts \cite{ouali2024clip}, contradicting the provided lyrics, as shown in Fig. 1. It's due to the auto-regressive nature of LLMs that amplifies minor prediction errors during sequential token generation, as well as the semantic gap between musical symbolism and linguistic representations in the latent space. However, relying solely on supervised fine-tuning (SFT) struggles to mitigate hallucination, because SFT's error accumulation in auto-regressive generation and the lack of corrective feedback. \par

\begin{figure}
    \centering
    \includegraphics[width=\linewidth]{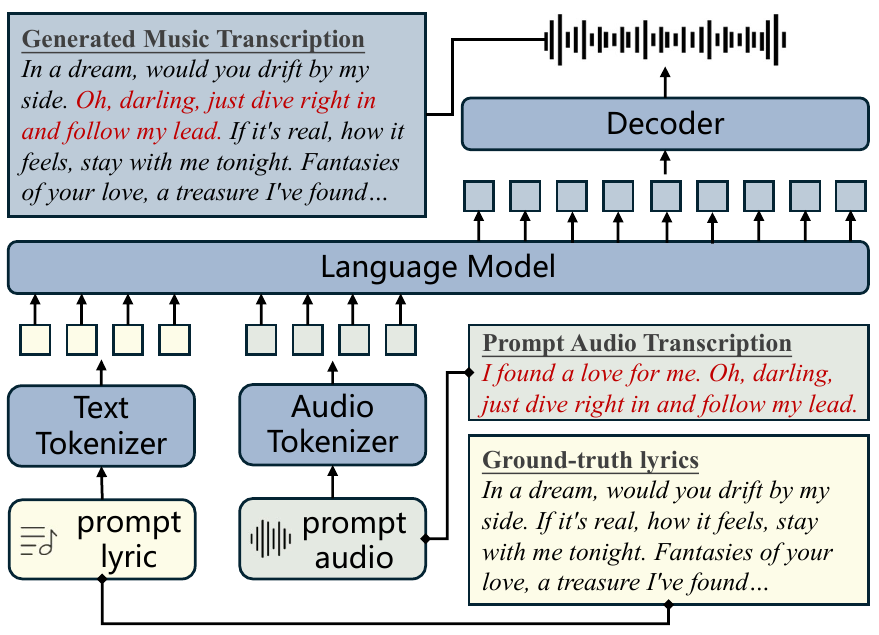}
    \caption{An example of hallucination in song generation.}
    % \label{fig:example} % 可选
\end{figure}

Fortunately, reinforcement learning (RL) has emerged as a powerful tool to align model performance with human preferences \cite{ziegler2019fine, zhang2024speechalign}. It has been proven effective in the alignment tasks of NLP \cite{yao2023deepspeed, shao2024deepseekmath, liu2025prorl, wang2025beyond}, computer vision \cite{zhang2024speechalign, sarkar2024mitigating}, audio \cite{anastassiou2024seed, cideron2024musicrl}, etc. Through its capacity for active preference optimization via reward shaping and direct intervention in the generation process, RL establishes a viable pathway for addressing and potentially eliminating hallucination phenomena in song generation. \par

In this work, we introduce a RL preference optimization framework that is tailored for song generation to mitigate hallucinations. The problem is first formulated as the reduction of the phoneme error rate (PER), which is utilized for constructing a preference dataset. Three tailored preference optimization strategies were specifically designed to mitigate music hallucinations, including off-policy Direct Preference Optimization (DPO), on-policy Proximal Preference Optimization (PPO), and on-policy Group Relative Policy Optimization (GRPO). After post-training, these methods effectively suppress anomalous generations with hallucinations in song generation. The main contributions are summarized in the following.
\begin{itemize}
\item  \textbf{Formalizing RL song hallucination mitigation}: We pioneer the formulation of song hallucination in song generation as a PER-driven preference optimization task. This provides a concrete, quantifiable framework for measuring and improving lyric-song alignment.
\item \textbf{Constructing a hallucination preference dataset}: We design a 3-step PER-guided filtering strategy to automatically construct chosen-rejected pairs. This dataset directly captures human-aligned preferences for hallucination mitigation, enabling effective off-policy optimization and reward model training.
\item \textbf{RL framework with multi-strategy optimization}: We develop and analyze a RL framework integrating three distinct preference optimization strategies (DPO, PPO, GRPO). This provides best practices for effective hallucination control in audio-based lyric-to-song generation.
\item \textbf{Transferable music RL paradigm}: we propose a versatile music RL framework, whose core design—adaptable preference data and reward models—enables seamless extension to diverse song generation tasks, offering a new paradigm for applying RL in song generation.
\end{itemize}

\section{Related Work}
\subsection{LLM-based Song Generation}
Early song generation research focuses on symbolic notation such as MIDI \cite{sturm2016music, wumidi}. Considering acquiring such notation is expensive and time-consuming, such methods are usually hard to achieve favorable performance. With the rapid development of LLMs, the next-token prediction becomes a new paradigm in audio generation tasks. The audios are first transferred to discrete audio tokens by neural codecs, whose representatives are SoundStream \cite{zeghidour2021soundstream} and Encodec \cite{wang2023neural}. Similar to text tokens, the audio tokens could be processed by LLMs for next-token prediction. In the domain of TTS, there exist many LLM-based methods that manifest remarkable performances with high naturalness and accuracy, such as AudioLM \cite{borsos2023audiolm}, seed-TTS \cite{anastassiou2024seed}, and VALL-E \cite{wang2023neural}. Following the demonstrated success, an increasing number of LLM-based technical iterations are now being actively explored for song generation tasks. Representatively, MusicLM casts the process of conditional song generation as a hierarchical sequence-to-sequence modeling task \cite{agostinelli2023musiclm}. SongGen introduces a single-stage auto-regressive transformer for controllable song generation \cite{liu2025songgen}. Seed-Music also leverages LLMs for controllable song generation; the performances can be controlled by multi-modal inputs \cite{bai2024seed}. Additionally, based on LLaMA2 \cite{touvron2023llama}, YuE is proposed to generate full-length songs with good lyric alignment and coherent music structures \cite{yuan2025yue}. Except for LLM-based methods, there are also some diffusion-based song generation methods like Ace-step \cite{gong2025ace} and DiffRhythm \cite{ning2025diffrhythm}.

\subsection{RL Post-training for Alignment}
Since LLMs are prone to amplifying minor prediction errors, and SFT fails to translate abstract human preferences into learnable gradient signals, RL is needed for alignment due to its exploration-exploitation mechanism. In the language domains, Deepspeed-Chat employs PPO \cite{schulman2017proximal} and builds an end-to-end Reinforcement Learning from Human Feedback (RLHF) language system \cite{yao2023deepspeed}. Based on PPO, VAPO \cite{yuan2025vapo} further explores tailored preference optimization in long sequence reasoning, especially Chain-of-Thought (CoT). In PPO and VAPO, a critic model is necessary to predict expected returns during token generation. The critic model makes the training more stable while bringing a high memory cost. Recently, there have been some critic-free preference optimization methods. GRPO proposed in Deepseek is one of the representatives, it eliminates the critic model through implicit reward modeling \cite{shao2024deepseekmath}. Moreover, DAPO \cite{yu2025dapo} integrates decoupled clip and dynamic sampling to optimize GRPO. Then, more variants of GRPO have existed for diverse applications \cite{zhang2025grpo, lin2025cppo}. From another perspective, DPO pioneers a reward-model-free paradigm for preference optimization, directly learning human preferences through policy-to-reference probability matching, thereby achieving more efficient style alignment \cite{rafailov2023direct}. In the vision domains, these RL technologies are also well applied, such as DPO in CLIP-DPO \cite{ouali2024clip}, PPO in DD-PPO \cite{wijmans2019dd}, and GRPO in GRPO-CARE \cite{chen2025grpo}.

\subsection{Preference Optimization in Audio}
RL-based preference optimization has emerged as a research hot-spot in audio processing. In TTS domains, RL preference optimization is employed to enhance speech quality \cite{anastassiou2024seed, zhang2024speechalign, chen2024enhancing}, according to human preference. It's utilized for alignment, which is for pursuing higher quality, narrowing distribution gaps, or enhancing zero-shot capabilities. Focusing on song generation, musicRL \cite{cideron2024musicrl} incorporates human feedback to enhance music quality and the overall subjective evaluation through RLHF, and seed-music \cite{bai2024seed} also introduces RL methods to enhance musicality and ensure alignment. \par

While some studies have successfully applied RL preference optimization to song generation, a dedicated method for addressing hallucination issues in large generative models through preference alignment is still lacking. Moreover, distinct preference optimization paradigms exhibit complementary strengths and limitations, making the exploration of optimal frameworks for musical hallucination mitigation an open research priority. Therefore, we have systematically explored multiple preference optimization methods and implemented targeted enhancements to control hallucinations in song generation.

\section{Method}

\subsection{Problem Formulation}
LLM hallucination describes a phenomenon in which the generated content appears nonsensical or unfaithful to the provided source content \cite{huang2025survey}. In song generation tasks, hallucination typically manifest as a mismatch between the generated content and the prompt, as shown in Fig. 1. In this work, an open-access song generation model, LeVo is utilized as the policy model to be optimized, whose audio LM is shown in Fig. 2. The input prompts are composed of a text prompt and an audio one, providing the lyrics and style, respectively. Encoded by the Byte Pair Encoding(BPE)-tokenizer and Codec Encoder, the token-like prompts can be used to generate song tokens. Finally, songs are generated by decoding such tokens with the Codec decoder. Nevertheless, the generated songs may deviate from the expected outcome—for instance, it might include phrases from the audio prompt like the situation in Fig. 1, exhibit inappropriate repetition inconsistent with the lyrics, or end abruptly before the lyrics are completely sung. Considering the success of RL methods in next-token-prediction tasks, RL preference optimization approaches can be employed to mitigate hallucinations in audio LMs.\par

\begin{figure}
    \centering
    \includegraphics[width=\linewidth]{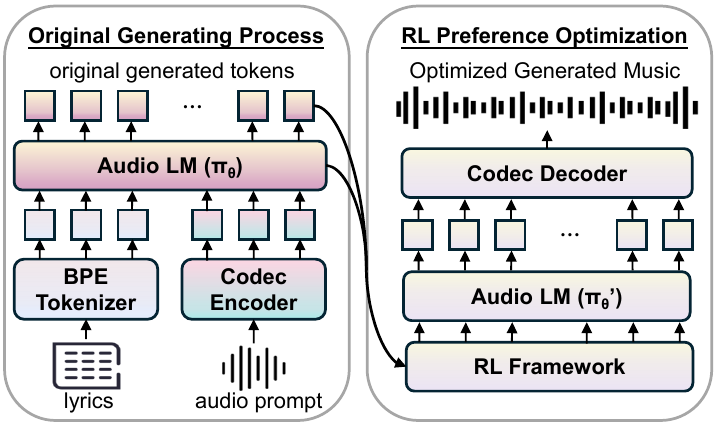}
    \caption{An audio language model generates songs sketches, and a RL framework is used as a preference optimization method to optimize the audio language model.}
    % \label{fig:example} % 可选
\end{figure}

Preference optimization requires organizing the generated data into a preference dataset that aligns with human preferences. To evaluate hallucinations, the mismatch between prompts and generated songs can be quantified as word (phoneme) error rate. The generated songs are processed by Automatic Speech Recognition (ASR) models, yielding recognized lyrics. However, since ASR models are prone to errors, and Chinese characters in songs often have ambiguous tonal pronunciations, the recognized characters usually differ significantly from the original lyrics. Therefore, they should be transferred to phonemes by Grapheme-to-Phoneme (G2P) models. Finally, hallucination mitigation issues are formulated as the reduction of PER. And the generated songs with PER can be utilized to compose a preference dataset, with details in Section 4.1.\par

\subsection{Reject Sampling}
With preference data, reject sampling (RS) is a simpler approach to optimizing models than RL methods. Specifically, only positive generated music is utilized for training in an SFT manner. Similar to the audio LM training, the positive samples are regarded as the target audios to be generated. Through RS, the model becomes more likely to generate tokens from positive samples, thereby learning the optimal alignment between lyrics and generated songs like positive examples. Thus, RS can be regarded as a fine-tuning method or the preliminary steps of off-policy DPO.

\begin{figure*}
    \centering
    \includegraphics[width=\linewidth]{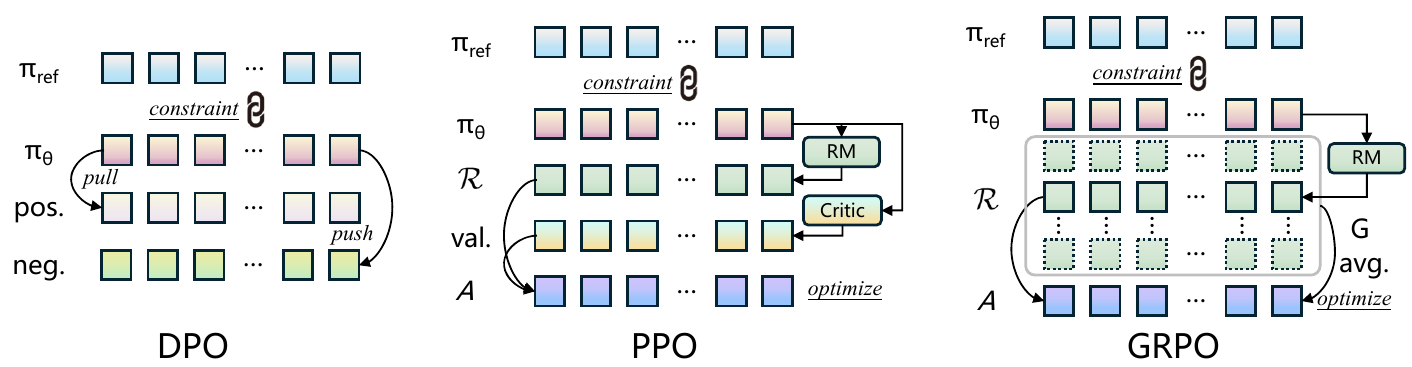}
    \caption{Illustrations of DPO, PPO, and GRPO in song generation hallucinations elimination. (val.: value, pos: positive samples, neg: negative sample, RM: reward model.)}
    % \label{fig:example} % 可选
\end{figure*}

\subsection{Song Hallucination DPO}
Since merely leveraging positive samples, RS suffers from low data efficiency and is prone to introducing noise. More effective RL methods are needed to better leverage preference data. DPO is a RL method that directly fine-tunes LMs using human preference data without explicit reward modeling \cite{rafailov2023direct}. Based on human preference data which composed of positive-negative pairs, DPO leverages the Bradley-Terry preference model to construct the loss function, directly maximizing the likelihood probability of preferred trajectories by contrasting positive and negative samples, thereby achieving more stable and efficient optimization, as shown in Fig. 3. Additionally, a reference model is employed to restrain the policy model, ensuring the policy model does not deviate too far from the original model. In Fig. 3, $\pi_\theta$ and $\pi_{ref}$ denote the policy and reference models, $y_w$ and $y_l$ stand for the positive and negative samples. The loss function is formulated as Eq. 1, where $\sigma$ is the activation function, and $\beta$ is a hyperparameter.

\begin{equation}
\begin{aligned}
\mathcal{L}_{DPO}(\pi _\theta ;\pi _{ref})=-\mathbb{E} _{(x,y_w,y_l)~\mathcal{D}} \\ 
\left [\log\sigma \left (\beta \log \frac{\pi _\theta(y_w|x)}{\pi _{ref}(y_w|x)}-\beta \log \frac{\pi _\theta(y_l|x)}{\pi _{ref}(y_l|x)} \right ) \right ]
\end{aligned}
\end{equation}

DPO training needs positive-negative pairs sourced from the same prompts; thereby, the generated songs with PER in the preference dataset should be correspondingly paired. To design the pairing rules, 360 generated songs sourced from 90 prompts with PER are evaluated and labeled as w/o hallucination by humans. Then, PER-based rules are made to filter out as much hallucinogenic data as possible. Finally, we propose a 3-step pairing strategy:
\begin{itemize}
    \item The difference in the number of incorrect phonemes between the two songs exceeds 40.
    \item If none of the four songs generated by a prompt form a pair, the songs with the highest and lowest PER are paired.
    \item Generated songs are classified as hallucinated data if they either insert $>$5 phonemes at one position or omit $>$10 phonemes within any 15-phoneme span compared to the original lyrics, with all hallucinated and non-hallucinated data paired.
\end{itemize}
The first rule focuses on the error length, because PER is related to the total song length. The length of the error segment can better reflect the degree of hallucination. And the second rule is employed for increasing the effective utilization of data. Even a group of generated songs does not create obvious hallucination, and the more accurate songs are promising to demonstrate a better routine. And the last rule is included to describe a mismatch from another perspective, especially for the hallucinations at the beginning and the abrupt ending at the conclusion. DPO achieves reinforcement learning based on model-generated data by maximizing the probability margin between the policy model generating preferred data and dispreferred data with the restraint of the reference model. \par

\subsection{Reward Model}
Since DPO is an off-policy approach, the generated songs are given a hidden reward by pairing. While on-policy methods do not prepare paired preference data in advance. To evaluate the generated songs, a reward model is necessary. In this work, the reward model has the same designation as the policy model, except for a regressive header attached after the final token. Such a position makes the whole sequence visible to the head. This design ensures consistent parameter spaces and representation power through shared architectures, improving training stability and initialization efficiency. In the hallucination mitigation task, we directly leverage the PER and define the reward score as Eq. 2, where $\mathcal{R}$ denotes the reward score. In other words, the reward model's regressive target becomes PER. L1 loss would be used to calculate the differences between the predicted rewards and the target calculated by PER.

\begin{equation}
\mathcal{R}=1-PER
\end{equation}

\subsection{Song Hallucination PPO}
Except for the reward model, PPO contains another 3 components: the policy, the reference, and the critic model \cite{yao2023deepspeed}. The policy model is the one to be reinforced, and the reference model provides constraints. And the reward model and the critic model verify the final and the expected return. PPO's action space is not sequence-level, but token-level. Each action is generating a specific token. Then the reward model and the critic model evaluate each action's true and expected returns. For stability, rewards for each token are formulated as Eq. 3 \cite{yao2023deepspeed}, where $t$ and $T$ denote the current and total time step, $p_{act}$ and $p_{ref}$ are the probability of the policy model and the reference model. $\alpha$ denotes the weight of the KL-divergence.
\begin{equation}
r(t) = 
\begin{cases} 
\alpha [\log p(\pi_{ref},t)-\log p(\pi_\theta,t)] \quad t \ne T\\
\alpha [\log p(\pi_{ref},t)-\log p(\pi_\theta,t)]+\mathcal{R}\quad t=T 
\end{cases}
\end{equation}

As for the prediction of expected returns, the critic model is trained to obtain a prediction value $V(t)$. With the rewards and expected returns, advantages could be computed. Maximizing the advantages means optimizing the actions with advantages, i.e., generating tokens that yield advantageous outcomes. To acquire such advantages, Temporal Difference (TD) error $\delta_t$ is introduced, which is formulated as Eq. 4. It represents how the returns beyond the expected.
\begin{equation}
\delta_t = r(t) + \gamma V(t+1) - V(t)
\end{equation}
Then, by accumulating the TD error, the advantages of the whole sequence can be computed by the General Advantage Estimation (GAE). As formulated in Eq. 5, $\lambda$ is a hyperparameter to balance the bias and variance.
\begin{equation}
A_t^{GAE}=\sum_{k=0}^{T-t}(\gamma\lambda)^k\delta_{t+k}
\end{equation}
However, song generation usually has massive tokens to be generated. Specifically, in this work, the generated 2-minute song would be transferred to 3000 tokens. Different from RL with fewer actions, the reward score would suffer a control loss if $\lambda$ is not equal to 1. For instance, if $\lambda$ is set as 0.99, then the reward score can only retain $0.99^{1000}$($4.32\times10^{-5}$) in the advantages for the 2000th token. For the first token, that value would be $0.99^{3000}$($8.05\times10^{-14}$). Therefore, we argue that the actual return of the complete trajectory is a better choice for long sequence generation. Thus, the GAE is transferred to Monte Carlo (MC) return, whose $\lambda$ is equal to 1. Finally, each token would be controlled by the reward score. Therefore, the policy loss and the critic loss are defined as Eq. 6-7. The policy loss contains an advantage loss and an entropy loss. Note that the KL divergence restraining the policy model by the reference model has already been incorporated in the reward function $r(t)$, i.e., it's incorporated in the advantage loss. The clip strategy in \cite{yao2023deepspeed} is retained to ensure the training stability. With respect to the entropy loss, it's employed for encouraging exploration, to alleviate explore-exploit puzzles.
\begin{equation}
\mathcal{L}_{pol} = \mathcal{L}_{adv} + \mathcal{L}_{entropy}, \mathcal{L}_{entropy} = -\mathbb{E}[\log p(a_t|s_t)]
\end{equation}
% \begin{equation}
% \mathcal{L}_{adv} = \mathbb{E}_t\left [ \min (c_tA_t^{GAE}, clip(c_t, 1\pm \epsilon)A_t^{GAE})\right ]
% \end{equation}

\begin{equation}
\mathcal{L}_{adv} = \mathbb{E}_t\left [clip(c_t, 1\pm \epsilon)A_t^{GAE}\right ], c_t = \frac{\log p(\pi_\theta)}{\log p(\pi_{\theta\_old)}}
\end{equation}

% \begin{equation}
% c_t = \frac{\log p_{act}}{\log p_{act\_old}}
% \end{equation}
% \begin{equation}
% \mathcal{L}_{entropy} = -\mathbb{E}[\log p(a_t|s_t)]
% \end{equation}

\subsection{Song Hallucination GRPO}
Although PPO is an effective method, it has a large computational cost and much hyperparameter to be elaborately tuned. In contrast, GRPO eliminates the critic model to reduce computational cost. Instead, the reward value after intra-group standardization is directly regarded as the advantage. The reward score is computed by Eq. (2). And for a particular group $G$, the advantage $A_i$ for sample $i$ is formulated as
\begin{equation}
A_i = \frac{\mathcal{R}_i-mean(\hat{\mathcal{R}}_G)}{std(\hat{\mathcal{R}}_G)}
\end{equation}
where mean and std stand for the average and the standard deviation of group $G$. By maximizing the advantages, the policy model is encouraged to generate more high-reward tokens. The loss function is formulated as Eq. (9), providing advantage optimization and the KL-divergence constraint. Note that a truncated sampling strategy is applied to screening high-entropy samples that can provide more typical information, accelerating and stabilizing the training process. It means that only the samples with the top-2 or bottom-2 reward score are used for training. 
\begin{equation}
\begin{aligned}
\mathcal{L}_{act} = -\frac{1}{4}\sum_{i=1}^{i\in I}A_i-\beta \mathcal{D}_{KL}(\pi_\theta, \pi_{ref})\\I=\{i\ if\ A_i\in top_2(A)\ or\ bottom_2(A) \}
\end{aligned}
\end{equation}
Furthermore, a K3 estimator is employed to ensure positivity KL-divergence \cite{shao2024deepseekmath}, as formulated in Eq. (10). Otherwise, the policy model could simply make the reference model output lower-probability tokens to artificially reduce the KL loss, which would lead to optimization failure. 
\begin{equation}
\mathcal{D}_{KL}(\pi_\theta, \pi_{ref}) = \frac{p(\pi_{ref},t)}{p(\pi_\theta,t)}-\log\frac{p(\pi_{ref},t)}{p(\pi_\theta, t)}+1
\end{equation}
Finally, the policy model could be optimized to alleviate hallucinations. However, properly trained song generation models tend to generate more hallucination-free samples. Thereby, the reward distribution is tightly clustered, with fewer low-reward outliers. Standardizing rewards using the $std$ may inadvertently suppress gradient signals from these critical low-reward samples—the primary targets for optimization. Additionally, an excessively small std might amplify noise in high-reward samples, leading to spurious negative gradients for correct tokens and greater gradient variance. Therefore, $std$ is removed in GRPO in this work; merely $\mathcal{R}_i-mean(\hat{\mathcal{R}}_G)$ is utilized as the advantage. 

\section{Experiments and Results}
\subsection{Experimental Setup}
\subsubsection{Preference Dataset Construction}
Preference optimization in generation models demands self-generated data. For song generation, text lyrics and prompt audio are necessary. In this work, Hunyuan-large \cite{sun2024hunyuan} is employed to generate 2857 lyrics with 10 styles. As for the prompt audio, 160 10-second song segments are acquired. Then, each prompt lyric is combined with randomly selected 10 prompt audio, yielding 21619 prompts. Since DPO requires paired data sharing the same prompts, each prompt is used to generate 4 songs. Finally, the dataset contains 86746 generated songs. As mentioned before, PER is regarded as the rule to compose preference data. To acquire PER, the vocal track of the generated songs is first extracted by Demucs \cite{defossez2019demucs}, then recognized and double-checked by Whisperlarge-v2 \cite{radford2023robust} and Zipformer \cite{yao2023zipformer}, and finally transferred to phonemes by G2P models. Finally, with the 3-step strategy in Section 3.3, these songs with PER compose 25459 paired preference data. 
\subsubsection{Subjective Evaluation Data}
For comprehensive evaluation, we designed a fixed set of subjective test prompts to generate audio samples for human listening evaluation. The prompts cover 10 distinct genres. In each experiment, models under evaluation would generate 20 songs covering these genres (2 songs per genre). The generated songs and genres are presented in the Demo.
\subsubsection{Evaluation Metrics}
In the experiments, the reward model is evaluated by the difference between predicted and recognized PER, which is calculated by L1 loss. And generated songs are evaluated by both objective and subjective experiments. For objective evaluation, Meta Audiobox-Aesthetic \cite{tjandra2025meta} is employed to score the aesthetic characteristics. The metrics are content enjoyment (CE), content usefulness (CU), production complexity (PC), and production quality (PQ).  In addition, a contrastive music-language pre-training model, MuQ-Mulan \cite{zhu2025muq} is employed to measure the similarity between generated songs and prompt audios. For subjective evaluation, each song receives ratings from 10 professional music annotators for the mean opinion score (MOS). The metrics are overall quality (OVL), vocal melodic attractiveness (MEL), vocal-instrument harmony (HAM), and lyrics following accuracy (LYC). More metric details are presented in the Appendix. 

\subsection{Reward Model Training}
To score generated songs for on-policy preference optimization, a reward model is trained by predicting the PER. The training is based on minimizing the difference between predicted and ground-truth PER. According to Fig. 4, with the validation loss converging to 0.03, the model achieves adequate performance to serve as a reliable reward model.
\begin{figure}
    \centering
    \includegraphics[width=0.9\linewidth]{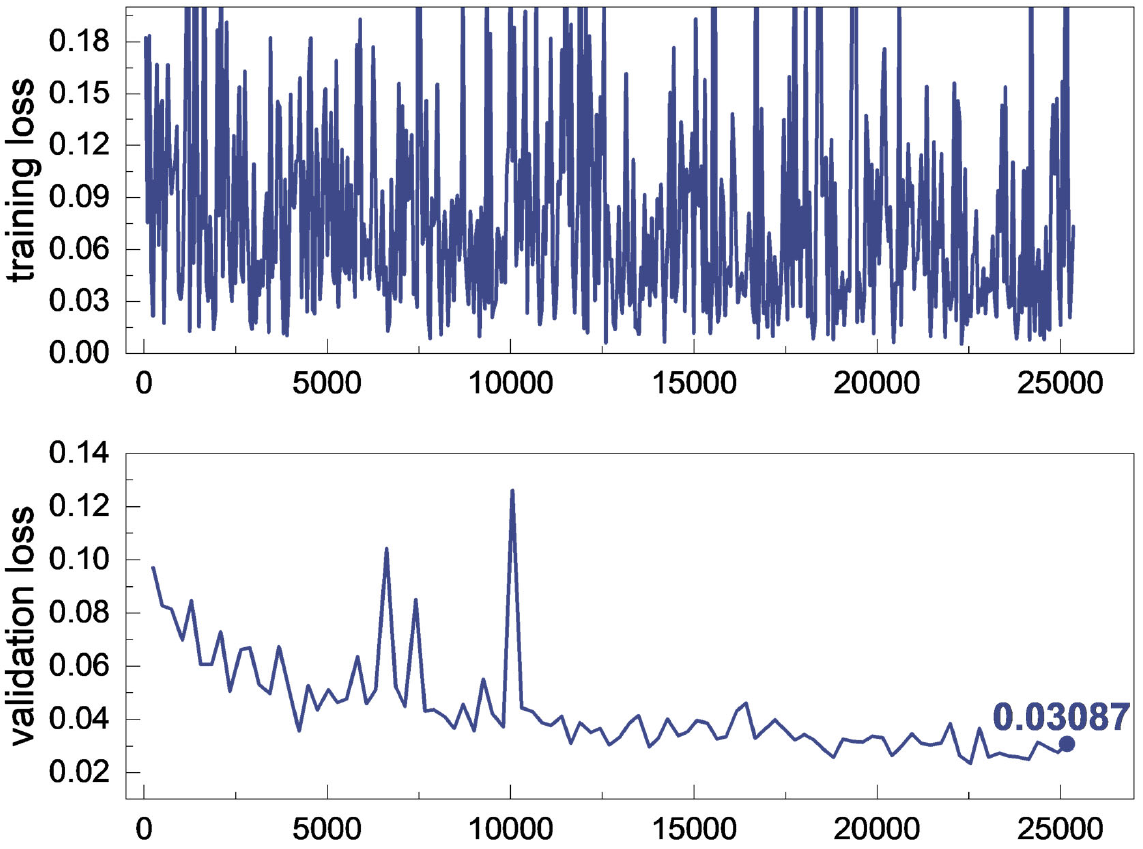}
    \caption{The training and validation loss of reward model.}
    % \label{fig:example} % 可选
\end{figure}

\subsection{Preference Optimization Training}
Using pre-collected preference data, DPO trains the model to favor tokens from positive (chosen) samples while avoiding those from negative (rejected) ones. The RS+DPO method is also evaluated, which is a dual optimizing method based on RS and DPO. According to the results illustrated in Fig. 5. (a), the probability gap between positive and negative tokens progressively widens during DPO training. It's worth mentioning that the probability curve of positive samples is decreasing in cases of RS+DPO. The rationale behind this is that RS has made the model more likely to generate positive tokens. Thus, during DPO training, the model corrects previous over-preference for specific tokens and shifts its optimization focus toward negative samples. Since DPO's targets are maximizing the probability margin between positive and negative tokens, DPO and RS+DPO exhibit distinct focuses on different portions of the preference data. \par 
On-policy methods, PPO and GRPO optimize the policy model by enlarging the advantages, which are acquired by on-policy generated sequences. According to Fig. 5. (b), a rise in the reward scores of generated sequences is observed, indicating progressive hallucinations mitigation. The non-monotonic reward trajectory suggests limitations of sparse reward design. Global sequence scoring fails to capture localized hallucination patterns, inducing misalignment between the scalar reward and segment-level quality. Therefore, an early stopping strategy was adopted to record the better-performing policy before model collapse occurs.

\subsection{Evaluating Hallucination Mitigation}
After preference optimization, the effectiveness of hallucination mitigation was evaluated in the experiments, as shown in Table 1. The average reward scores $\hat{r}$ of songs generated from validation prompts are presented. Empirically, samples with reward scores $<$0.7 ($>$0.8) typically exhibit whether they are with (w/o) hallucinations. Due to errors in ASR and G2P processing, the correlation between reward scores (0.7–0.8) and hallucination occurrences is non-deterministic. For intuitive demonstration, we define 0.7 and 0.8 as the deterministic thresholds for hallucinated and hallucination-free outputs. \par 
According to the results, all the methods within the proposed framework demonstrate superior hallucination mitigation compared to the baseline model. The best method is DPO with RS, which elevates reward scores by 9.60\%, concurrently reducing hallucinated samples by 16.5\% while increasing hallucination-free outputs by 38.47\%. Therefore, the proposed framework fundamentally resolves hallucination issues in song generation.

\begin{figure}
    \centering
    \includegraphics[width=\linewidth]{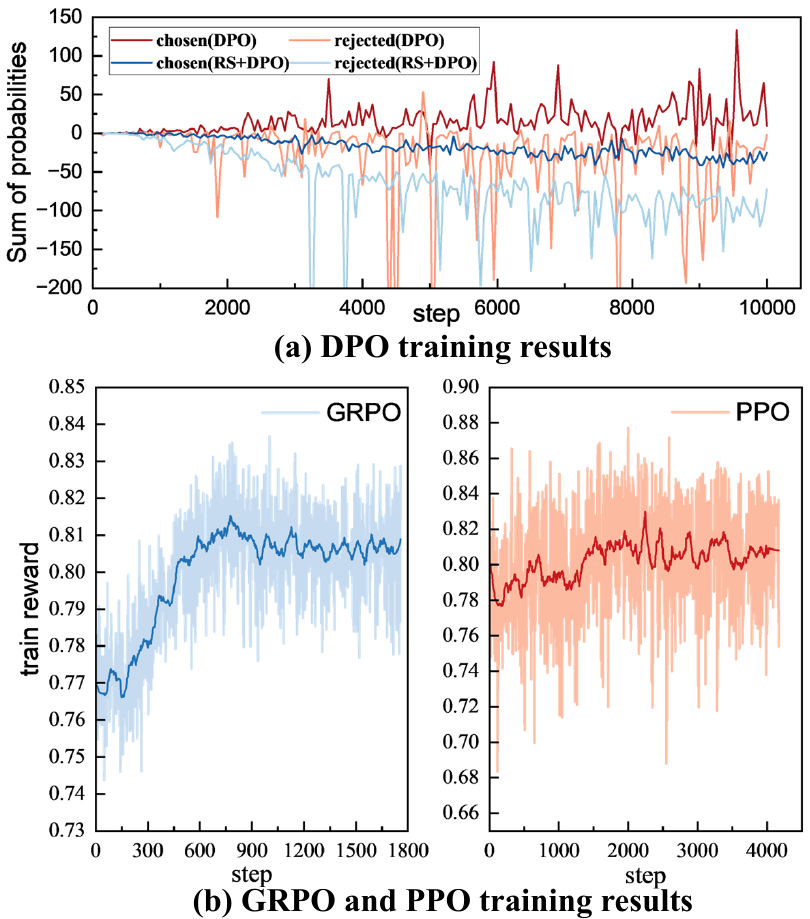}
    \caption{Results of RL training. (a) The sum of token probabilities in the chosen/rejected samples generated by the policy model in DPO. (b) Reward scores of songs generated by the policy model during GRPO and PPO training.}
    % \label{fig:example} % 可选
\end{figure}
\begin{table}[htbp]
  \centering
  % \caption{Reward score and reward distribution for generated songs based on validation prompts. $PER=1-r$.}
    \begin{tabularx}{\linewidth}{c|cccc}  % X列自动分配剩余宽度
    \toprule
    Method & $\hat{r}\uparrow$ & $r<0.7 \downarrow$ & 0.7-0.8 & $r>0.8\uparrow$ \\
    \midrule
    Origin & 0.771  & 17.30\% & 39.44\% & 43.26\% \\
    RS    & 0.783  & 16.34\% & 29.81\% & 53.85\% \\
    DPO   & 0.843  & 4.98\% & 14.17\% & 80.85\% \\
    RS+DPO & \textbf{0.845} & \textbf{4.80\%} & 13.47\% & \textbf{81.73\%} \\
    PPO   & 0.820  & 6.73\% & 23.06\% & 70.20\% \\
    GRPO  & 0.834  & 6.73\% & 17.30\% & 75.97\% \\
    \bottomrule
    \end{tabularx}%
  \caption{Reward score and reward distribution for generated songs based on validation prompts. $PER=1-r$.}
  \label{tab1}%
\end{table}

\subsection{Objective Content Evaluation}
While preference optimization has substantially mitigated hallucination issues, it is equally critical to ensure that musicality and audio quality are not compromised during this process. Results are shown in Table 2. The aesthetic results of all the methods are nearly the same. Empirically, Audiobox-aesthetic scores require a minimum difference of 0.1 points to reliably discern qualitative superiority. Based on such judgment, all methods demonstrate marginal differences within 0.1 on CE, CU, and PQ, while preference-optimized methods exhibit a maximum lead of 0.18 on PC. Moreover, the similar performances on MuQ confirm that the audio prompt conditions remain unaffected throughout the process. Therefore, the proposed framework achieves hallucination mitigation while preserving musical aesthetics.

\begin{table}[htbp]
  \centering
  % \caption{Objective results of the baseline model vs. the preference-optimized models. The overall first and second results are marked with \textbf{bold} and \underline{underline}, respectively.}
    \begin{tabularx}{\linewidth}{@{}l*{5}{>{\centering\arraybackslash}X}@{}}
    \toprule
    \multicolumn{1}{c}{\textbf{Methods}} & \textbf{MuQ}$\uparrow$ & \textbf{CE}$\uparrow$ & \textbf{CU}$\uparrow$ & \textbf{PC}$\uparrow$ & \textbf{PQ}$\uparrow$ \\
    \midrule
    Origin & 0.80 & \textbf{7.63} & \textbf{7.85} & 6.35  & \underline{8.40}  \\
    RS    & 0.79 & \textbf{7.63} & 7.83  & \textbf{6.54} & 8.39  \\
    DPO   & \underline{0.81} & \textbf{7.63} & 7.84  & 6.42  & 8.39  \\
    DPO+RS & 0.80 & \textbf{7.63} & \textbf{7.85} & 6.48  & \textbf{8.42} \\
    PPO   & 0.78 & 7.58  & 7.78  & \underline{6.53}  & 8.31  \\
    GRPO  & \textbf{0.82}& 7.58  & 7.80  & 6.51  & 8.36  \\
    \bottomrule
    \end{tabularx}%
  \caption{Objective results of the baseline model vs. the preference-optimized models. The overall first and second results are marked with \textbf{bold} and \underline{underline}, respectively.}
  \label{tab2}%
\end{table}%

\subsection{Subjective Evaluation}
To ensure evaluation alignment with human preferences, subjective experiments are conducted, as shown in Table 3. It reveals that musicality metrics (OVL, MEL, HAM) are not compromised despite hallucination mitigation. LYC is a metric for hallucination in subjective evaluation. LYC results demonstrate varying degrees of hallucination mitigation enhancement across different preference optimization approaches. Consistent with reward score experiments, DPO with RS presents the best hallucination mitigation performance. To sum up, preference optimization successfully aligns generated songs with provided lyrics without a decrease in musicality and audio quality. The complete results with 95\% confidence intervals (CI) are presented in the Appendix.
% Table generated by Excel2LaTeX from sheet 'Sheet2'
\begin{table}[htbp]
  \centering
  % \caption{Subjective results of the baseline model vs. the preference-optimized models. The overall first and second results are marked with \textbf{bold} and \underline{underline}, respectively.}
    \begin{tabular}{l|ccccccc}
    \toprule
    \multicolumn{1}{c|}{\multirow{2}[4]{*}{Methods}} & \multicolumn{7}{c}{\textbf{MOS}$\uparrow$} \\
\cmidrule{2-2}\cmidrule{4-4}\cmidrule{6-6}\cmidrule{8-8}          & \textbf{OVL} &       & \textbf{MEL} &       & \textbf{HAM} &       & \textbf{LYC} \\
    \midrule
    Origin & \underline{3.040}  &       & 3.455  &       & 3.595  &       & 2.605  \\
    RS    & \textbf{3.070} &       & \textbf{3.705}     &       & \underline{3.615}     &       & \underline{2.930} \\
    DPO   & 3.020  &       & 3.525  &       & \textbf{3.650}  &       & 2.850  \\
    RS+DPO & \underline{3.040}  &       & \underline{3.560}  &       & 3.530  &       & \textbf{3.160}  \\
    PPO   & 2.965  &       & 3.330  &       & 3.605  &       & 2.790  \\
    GRPO  & 2.970  &       & 3.430  &       & 3.460  &       & 2.180  \\
    \bottomrule
    \end{tabular}%
  \caption{Subjective results of the baseline model vs. the preference-optimized models. The overall first and second results are marked with \textbf{bold} and \underline{underline}, respectively.}
  \label{tab3}%
\end{table}%

\section{Discussion and Conclusion}
\subsection{Method Transferability}
Although song hallucination is the main optimization object in this work, the proposed framework can be easily transferred to other song generation tasks. For example, generated songs can be annotated as pairs or ranks to construct preference datasets for DPO with RS. Additionally, reward models can be trained based on generated songs with scoring. Thereby, PPO and GRPO can be directly applied. The preference optimization methods within the framework are suitable for different kinds of tasks, focusing on general, local, or semantic information. Optimal strategies can be chosen for tasks with specific characteristics, presenting good transferability of the proposed framework.

\subsection{Limitations and Future Research}
Despite the effectiveness of hallucination mitigation, the proposed framework still has some problems. Theoretically, on-policy methods exhibit superior performance ceilings over off-policy ones, as the sampled data consistently reflects the most current policy performance. However, it contradicts the experimental results observed. This primarily stems from the sparsity of using a single reward score to represent the entire sequence, which fails to strongly correlate with hallucination locations. Consequently, the training process misjudges numerous normal tokens. In the future, the reward models need more complex designs. Training data also requires more granular annotation. Moreover, the proposed framework will be applied to other song generation tasks, reinforcing the music LM from different aspects, like musicality and quality.

\subsection{Conclusion}
This work presents a novel RL framework for reliable, hallucination-controlled lyric-to-song generation. By formulating lyric alignment as a PER-driven preference optimization task, we establish the first quantifiable approach to mitigating music hallucination. Key to our solution is the construction of a hallucination-aware preference dataset through PER-guided filtering. Within the proposed music RL paradigm, we implement and evaluate three preference optimization strategies—DPO, PPO, and GRPO—demonstrating their efficacy in hallucination mitigation. Comprehensive experiments confirm that our framework significantly reduces misalignment (e.g., DPO with RS achieves a 16.50\% reduction in hallucinated songs and a 38.47\% increase in hallucination-free samples, with a 9.60\% reward score gain) while preserving musicality and audio quality. Crucially, the transferable design of our framework—centered on adaptable preference data and reward models—readily extends to diverse song generation challenges, such as musicality enhancement and style control. This work establishes preference-based RL as a foundational paradigm for achieving high-fidelity, human-aligned generative song systems.

\bibliography{references}

\begin{thebibliography}{42}
\providecommand{\natexlab}[1]{#1}

\bibitem[{Achiam et~al.(2023)Achiam, Adler, Agarwal, Ahmad, Akkaya, Aleman,
  Almeida, Altenschmidt, Altman, Anadkat et~al.}]{achiam2023gpt}
Achiam, J.; Adler, S.; Agarwal, S.; Ahmad, L.; Akkaya, I.; Aleman, F.~L.;
  Almeida, D.; Altenschmidt, J.; Altman, S.; Anadkat, S.; et~al. 2023.
\newblock Gpt-4 technical report.
\newblock \emph{arXiv preprint arXiv:2303.08774}.

\bibitem[{Agostinelli et~al.(2023)Agostinelli, Denk, Borsos, Engel, Verzetti,
  Caillon, Huang, Jansen, Roberts, Tagliasacchi
  et~al.}]{agostinelli2023musiclm}
Agostinelli, A.; Denk, T.~I.; Borsos, Z.; Engel, J.; Verzetti, M.; Caillon, A.;
  Huang, Q.; Jansen, A.; Roberts, A.; Tagliasacchi, M.; et~al. 2023.
\newblock Musiclm: Generating music from text.
\newblock \emph{arXiv preprint arXiv:2301.11325}.

\bibitem[{Anastassiou et~al.(2024)Anastassiou, Chen, Chen, Chen, Chen, Chen,
  Cong, Deng, Ding, Gao et~al.}]{anastassiou2024seed}
Anastassiou, P.; Chen, J.; Chen, J.; Chen, Y.; Chen, Z.; Chen, Z.; Cong, J.;
  Deng, L.; Ding, C.; Gao, L.; et~al. 2024.
\newblock Seed-tts: A family of high-quality versatile speech generation
  models.
\newblock \emph{arXiv preprint arXiv:2406.02430}.

\bibitem[{Bai et~al.(2024)Bai, Chen, Chen, Chen, Deng, Dong, Hantrakul, Hao,
  Huang, Huang et~al.}]{bai2024seed}
Bai, Y.; Chen, H.; Chen, J.; Chen, Z.; Deng, Y.; Dong, X.; Hantrakul, L.; Hao,
  W.; Huang, Q.; Huang, Z.; et~al. 2024.
\newblock Seed-music: A unified framework for high quality and controlled music
  generation.
\newblock \emph{arXiv preprint arXiv:2409.09214}.

\bibitem[{Borsos et~al.(2023)Borsos, Marinier, Vincent, Kharitonov, Pietquin,
  Sharifi, Roblek, Teboul, Grangier, Tagliasacchi et~al.}]{borsos2023audiolm}
Borsos, Z.; Marinier, R.; Vincent, D.; Kharitonov, E.; Pietquin, O.; Sharifi,
  M.; Roblek, D.; Teboul, O.; Grangier, D.; Tagliasacchi, M.; et~al. 2023.
\newblock Audiolm: a language modeling approach to audio generation.
\newblock \emph{IEEE/ACM transactions on audio, speech, and language
  processing}, 31: 2523--2533.

\bibitem[{Chen et~al.(2024)Chen, Hu, Wu, Wang, Chng, and
  Zhang}]{chen2024enhancing}
Chen, C.; Hu, Y.; Wu, W.; Wang, H.; Chng, E.~S.; and Zhang, C. 2024.
\newblock Enhancing zero-shot text-to-speech synthesis with human feedback.
\newblock \emph{arXiv preprint arXiv:2406.00654}.

\bibitem[{Chen et~al.(2025)Chen, Ge, Wang, Ge, Cheng, Shan, and
  Liu}]{chen2025grpo}
Chen, Y.; Ge, Y.; Wang, R.; Ge, Y.; Cheng, J.; Shan, Y.; and Liu, X. 2025.
\newblock GRPO-CARE: Consistency-Aware Reinforcement Learning for Multimodal
  Reasoning.
\newblock \emph{arXiv preprint arXiv:2506.16141}.

\bibitem[{Cideron et~al.(2024)Cideron, Girgin, Verzetti, Vincent, Kastelic,
  Borsos, Mcwilliams, Ungureanu, Bachem, Pietquin et~al.}]{cideron2024musicrl}
Cideron, G.; Girgin, S.; Verzetti, M.; Vincent, D.; Kastelic, M.; Borsos, Z.;
  Mcwilliams, B.; Ungureanu, V.; Bachem, O.; Pietquin, O.; et~al. 2024.
\newblock MusicRL: Aligning Music Generation to Human Preferences.
\newblock In \emph{International Conference on Machine Learning}, 8968--8984.
  PMLR.

\bibitem[{D{\'e}fossez et~al.(2019)D{\'e}fossez, Usunier, Bottou, and
  Bach}]{defossez2019demucs}
D{\'e}fossez, A.; Usunier, N.; Bottou, L.; and Bach, F. 2019.
\newblock Demucs: Deep extractor for music sources with extra unlabeled data
  remixed.
\newblock \emph{arXiv preprint arXiv:1909.01174}.

\bibitem[{Du et~al.(2024)Du, Wang, Chen, Shi, Lv, Zhao, Gao, Yang, Gao, Wang
  et~al.}]{du2024cosyvoice}
Du, Z.; Wang, Y.; Chen, Q.; Shi, X.; Lv, X.; Zhao, T.; Gao, Z.; Yang, Y.; Gao,
  C.; Wang, H.; et~al. 2024.
\newblock Cosyvoice 2: Scalable streaming speech synthesis with large language
  models.
\newblock \emph{arXiv preprint arXiv:2412.10117}.

\bibitem[{Gong et~al.(2025)Gong, Zhao, Wang, Xu, and Guo}]{gong2025ace}
Gong, J.; Zhao, S.; Wang, S.; Xu, S.; and Guo, J. 2025.
\newblock Ace-step: A step towards music generation foundation model.
\newblock \emph{arXiv preprint arXiv:2506.00045}.

\bibitem[{Huang et~al.(2025)Huang, Yu, Ma, Zhong, Feng, Wang, Chen, Peng, Feng,
  Qin et~al.}]{huang2025survey}
Huang, L.; Yu, W.; Ma, W.; Zhong, W.; Feng, Z.; Wang, H.; Chen, Q.; Peng, W.;
  Feng, X.; Qin, B.; et~al. 2025.
\newblock A survey on hallucination in large language models: Principles,
  taxonomy, challenges, and open questions.
\newblock \emph{ACM Transactions on Information Systems}, 43(2): 1--55.

\bibitem[{Ju et~al.(2024)Ju, Wang, Shen, Tan, Xin, Yang, Liu, Leng, Song, Tang
  et~al.}]{ju2024naturalspeech}
Ju, Z.; Wang, Y.; Shen, K.; Tan, X.; Xin, D.; Yang, D.; Liu, E.; Leng, Y.;
  Song, K.; Tang, S.; et~al. 2024.
\newblock NaturalSpeech 3: Zero-Shot Speech Synthesis with Factorized Codec and
  Diffusion Models.
\newblock In \emph{International Conference on Machine Learning}, 22605--22623.
  PMLR.

\bibitem[{Lei et~al.(2025)Lei, Xu, Lin, Zhang, Tan, Chen, Yu, Zhang, Yang, Zhu
  et~al.}]{lei2025levo}
Lei, S.; Xu, Y.; Lin, Z.; Zhang, H.; Tan, W.; Chen, H.; Yu, J.; Zhang, Y.;
  Yang, C.; Zhu, H.; et~al. 2025.
\newblock LeVo: High-Quality Song Generation with Multi-Preference Alignment.
\newblock \emph{arXiv preprint arXiv:2506.07520}.

\bibitem[{Lin et~al.(2025)Lin, Lin, Xie, and Ji}]{lin2025cppo}
Lin, Z.; Lin, M.; Xie, Y.; and Ji, R. 2025.
\newblock Cppo: Accelerating the training of group relative policy
  optimization-based reasoning models.
\newblock \emph{arXiv preprint arXiv:2503.22342}.

\bibitem[{Liu et~al.(2025{\natexlab{a}})Liu, Diao, Lu, Hu, Dong, Choi, Kautz,
  and Dong}]{liu2025prorl}
Liu, M.; Diao, S.; Lu, X.; Hu, J.; Dong, X.; Choi, Y.; Kautz, J.; and Dong, Y.
  2025{\natexlab{a}}.
\newblock Prorl: Prolonged reinforcement learning expands reasoning boundaries
  in large language models.
\newblock \emph{arXiv preprint arXiv:2505.24864}.

\bibitem[{Liu et~al.(2025{\natexlab{b}})Liu, Ding, Zhang, Dong, Zhang, Zang,
  Cao, Lin, and Wang}]{liu2025songgen}
Liu, Z.; Ding, S.; Zhang, Z.; Dong, X.; Zhang, P.; Zang, Y.; Cao, Y.; Lin, D.;
  and Wang, J. 2025{\natexlab{b}}.
\newblock Songgen: A single stage auto-regressive transformer for text-to-song
  generation.
\newblock \emph{arXiv preprint arXiv:2502.13128}.

\bibitem[{Ning et~al.(2025)Ning, Chen, Jiang, Hao, Ma, Wang, Yao, and
  Xie}]{ning2025diffrhythm}
Ning, Z.; Chen, H.; Jiang, Y.; Hao, C.; Ma, G.; Wang, S.; Yao, J.; and Xie, L.
  2025.
\newblock DiffRhythm: Blazingly fast and embarrassingly simple end-to-end
  full-length song generation with latent diffusion.
\newblock \emph{arXiv preprint arXiv:2503.01183}.

\bibitem[{Ouali et~al.(2024)Ouali, Bulat, Martinez, and
  Tzimiropoulos}]{ouali2024clip}
Ouali, Y.; Bulat, A.; Martinez, B.; and Tzimiropoulos, G. 2024.
\newblock Clip-dpo: Vision-language models as a source of preference for fixing
  hallucinations in lvlms.
\newblock In \emph{European Conference on Computer Vision}, 395--413. Springer.

\bibitem[{Radford et~al.(2023)Radford, Kim, Xu, Brockman, McLeavey, and
  Sutskever}]{radford2023robust}
Radford, A.; Kim, J.~W.; Xu, T.; Brockman, G.; McLeavey, C.; and Sutskever, I.
  2023.
\newblock Robust speech recognition via large-scale weak supervision.
\newblock In \emph{International conference on machine learning}, 28492--28518.
  PMLR.

\bibitem[{Rafailov et~al.(2023)Rafailov, Sharma, Mitchell, Manning, Ermon, and
  Finn}]{rafailov2023direct}
Rafailov, R.; Sharma, A.; Mitchell, E.; Manning, C.~D.; Ermon, S.; and Finn, C.
  2023.
\newblock Direct preference optimization: Your language model is secretly a
  reward model.
\newblock \emph{Advances in Neural Information Processing Systems}, 36:
  53728--53741.

\bibitem[{Sarkar et~al.(2024)Sarkar, Ebrahimi, Etemad, Beirami, Ar{\i}k, and
  Pfister}]{sarkar2024mitigating}
Sarkar, P.; Ebrahimi, S.; Etemad, A.; Beirami, A.; Ar{\i}k, S.~{\"O}.; and
  Pfister, T. 2024.
\newblock Mitigating Object Hallucination in MLLMs via Data-augmented
  Phrase-level Alignment.
\newblock \emph{arXiv preprint arXiv:2405.18654}.

\bibitem[{Schulman et~al.(2017)Schulman, Wolski, Dhariwal, Radford, and
  Klimov}]{schulman2017proximal}
Schulman, J.; Wolski, F.; Dhariwal, P.; Radford, A.; and Klimov, O. 2017.
\newblock Proximal policy optimization algorithms.
\newblock \emph{arXiv preprint arXiv:1707.06347}.

\bibitem[{Shao et~al.(2024)Shao, Wang, Zhu, Xu, Song, Bi, Zhang, Zhang, Li, Wu
  et~al.}]{shao2024deepseekmath}
Shao, Z.; Wang, P.; Zhu, Q.; Xu, R.; Song, J.; Bi, X.; Zhang, H.; Zhang, M.;
  Li, Y.; Wu, Y.; et~al. 2024.
\newblock Deepseekmath: Pushing the limits of mathematical reasoning in open
  language models.
\newblock \emph{arXiv preprint arXiv:2402.03300}.

\bibitem[{Sturm et~al.(2016)Sturm, Santos, Ben-Tal, and
  Korshunova}]{sturm2016music}
Sturm, B.; Santos, J.~F.; Ben-Tal, O.; and Korshunova, I. 2016.
\newblock Music Transcription Modelling and Composition Using Deep Learning.
\newblock In \emph{1st Conference on Computer Simulation of Musical
  Creativity}.

\bibitem[{Sun et~al.(2024)Sun, Chen, Huang, Xie, Zhu, Zhang, Li, Yang, Han, Shu
  et~al.}]{sun2024hunyuan}
Sun, X.; Chen, Y.; Huang, Y.; Xie, R.; Zhu, J.; Zhang, K.; Li, S.; Yang, Z.;
  Han, J.; Shu, X.; et~al. 2024.
\newblock Hunyuan-large: An open-source moe model with 52 billion activated
  parameters by tencent.
\newblock \emph{arXiv preprint arXiv:2411.02265}.

\bibitem[{Tjandra et~al.(2025)Tjandra, Wu, Guo, Hoffman, Ellis, Vyas, Shi,
  Chen, Le, Zacharov et~al.}]{tjandra2025meta}
Tjandra, A.; Wu, Y.-C.; Guo, B.; Hoffman, J.; Ellis, B.; Vyas, A.; Shi, B.;
  Chen, S.; Le, M.; Zacharov, N.; et~al. 2025.
\newblock Meta audiobox aesthetics: Unified automatic quality assessment for
  speech, music, and sound.
\newblock \emph{arXiv preprint arXiv:2502.05139}.

\bibitem[{Touvron et~al.(2023)Touvron, Martin, Stone, Albert, Almahairi,
  Babaei, Bashlykov, Batra, Bhargava, Bhosale et~al.}]{touvron2023llama}
Touvron, H.; Martin, L.; Stone, K.; Albert, P.; Almahairi, A.; Babaei, Y.;
  Bashlykov, N.; Batra, S.; Bhargava, P.; Bhosale, S.; et~al. 2023.
\newblock Llama 2: Open foundation and fine-tuned chat models.
\newblock \emph{arXiv preprint arXiv:2307.09288}.

\bibitem[{Wang et~al.(2023)Wang, Chen, Wu, Zhang, Zhou, Liu, Chen, Liu, Wang,
  Li et~al.}]{wang2023neural}
Wang, C.; Chen, S.; Wu, Y.; Zhang, Z.; Zhou, L.; Liu, S.; Chen, Z.; Liu, Y.;
  Wang, H.; Li, J.; et~al. 2023.
\newblock Neural codec language models are zero-shot text to speech
  synthesizers.
\newblock \emph{arXiv preprint arXiv:2301.02111}.

\bibitem[{Wang et~al.(2025)Wang, Yu, Gao, Zheng, Liu, Lu, Dang, Chen, Yang,
  Zhang et~al.}]{wang2025beyond}
Wang, S.; Yu, L.; Gao, C.; Zheng, C.; Liu, S.; Lu, R.; Dang, K.; Chen, X.;
  Yang, J.; Zhang, Z.; et~al. 2025.
\newblock Beyond the 80/20 rule: High-entropy minority tokens drive effective
  reinforcement learning for llm reasoning.
\newblock \emph{arXiv preprint arXiv:2506.01939}.

\bibitem[{Wijmans et~al.(2019)Wijmans, Kadian, Morcos, Lee, Essa, Parikh,
  Savva, and Batra}]{wijmans2019dd}
Wijmans, E.; Kadian, A.; Morcos, A.; Lee, S.; Essa, I.; Parikh, D.; Savva, M.;
  and Batra, D. 2019.
\newblock Dd-ppo: Learning near-perfect pointgoal navigators from 2.5 billion
  frames.
\newblock \emph{arXiv preprint arXiv:1911.00357}.

\bibitem[{Wu et~al.(2022)Wu, Manilow, Deng, Swavely, Kastner, Cooijmans,
  Courville, Huang, and Engel}]{wumidi}
Wu, Y.; Manilow, E.; Deng, Y.; Swavely, R.; Kastner, K.; Cooijmans, T.;
  Courville, A.; Huang, C.-Z.~A.; and Engel, J. 2022.
\newblock MIDI-DDSP: Detailed Control of Musical Performance via Hierarchical
  Modeling.
\newblock In \emph{International Conference on Learning Representations}.

\bibitem[{Yao et~al.(2023{\natexlab{a}})Yao, Aminabadi, Ruwase, Rajbhandari,
  Wu, Awan, Rasley, Zhang, Li, Holmes et~al.}]{yao2023deepspeed}
Yao, Z.; Aminabadi, R.~Y.; Ruwase, O.; Rajbhandari, S.; Wu, X.; Awan, A.~A.;
  Rasley, J.; Zhang, M.; Li, C.; Holmes, C.; et~al. 2023{\natexlab{a}}.
\newblock Deepspeed-chat: Easy, fast and affordable rlhf training of
  chatgpt-like models at all scales.
\newblock \emph{arXiv preprint arXiv:2308.01320}.

\bibitem[{Yao et~al.(2023{\natexlab{b}})Yao, Guo, Yang, Kang, Kuang, Yang, Jin,
  Lin, and Povey}]{yao2023zipformer}
Yao, Z.; Guo, L.; Yang, X.; Kang, W.; Kuang, F.; Yang, Y.; Jin, Z.; Lin, L.;
  and Povey, D. 2023{\natexlab{b}}.
\newblock Zipformer: A faster and better encoder for automatic speech
  recognition.
\newblock \emph{arXiv preprint arXiv:2310.11230}.

\bibitem[{Yu et~al.(2025)Yu, Zhang, Zhu, Yuan, Zuo, Yue, Fan, Liu, Liu, Liu
  et~al.}]{yu2025dapo}
Yu, Q.; Zhang, Z.; Zhu, R.; Yuan, Y.; Zuo, X.; Yue, Y.; Fan, T.; Liu, G.; Liu,
  L.; Liu, X.; et~al. 2025.
\newblock Dapo: An open-source llm reinforcement learning system at scale.
\newblock \emph{arXiv preprint arXiv:2503.14476}.

\bibitem[{Yuan et~al.(2025{\natexlab{a}})Yuan, Lin, Guo, Zhang, Pan, Zang, Liu,
  Liang, Ma, Du et~al.}]{yuan2025yue}
Yuan, R.; Lin, H.; Guo, S.; Zhang, G.; Pan, J.; Zang, Y.; Liu, H.; Liang, Y.;
  Ma, W.; Du, X.; et~al. 2025{\natexlab{a}}.
\newblock YuE: Scaling Open Foundation Models for Long-Form Music Generation.
\newblock \emph{arXiv preprint arXiv:2503.08638}.

\bibitem[{Yuan et~al.(2025{\natexlab{b}})Yuan, Yu, Zuo, Zhu, Xu, Chen, Wang,
  Fan, Du, Wei et~al.}]{yuan2025vapo}
Yuan, Y.; Yu, Q.; Zuo, X.; Zhu, R.; Xu, W.; Chen, J.; Wang, C.; Fan, T.; Du,
  Z.; Wei, X.; et~al. 2025{\natexlab{b}}.
\newblock VAPO: Efficient and reliable reinforcement learning for advanced
  reasoning tasks.
\newblock \emph{arXiv preprint arXiv:2504.05118}.

\bibitem[{Zeghidour et~al.(2021)Zeghidour, Luebs, Omran, Skoglund, and
  Tagliasacchi}]{zeghidour2021soundstream}
Zeghidour, N.; Luebs, A.; Omran, A.; Skoglund, J.; and Tagliasacchi, M. 2021.
\newblock Soundstream: An end-to-end neural audio codec.
\newblock \emph{IEEE/ACM Transactions on Audio, Speech, and Language
  Processing}, 30: 495--507.

\bibitem[{Zhang et~al.(2024)Zhang, Li, Li, Zhang, Wang, Zhou, and
  Qiu}]{zhang2024speechalign}
Zhang, D.; Li, Z.; Li, S.; Zhang, X.; Wang, P.; Zhou, Y.; and Qiu, X. 2024.
\newblock Speechalign: Aligning speech generation to human preferences.
\newblock \emph{Advances in Neural Information Processing Systems}, 37:
  50343--50360.

\bibitem[{Zhang and Zuo(2025)}]{zhang2025grpo}
Zhang, J.; and Zuo, C. 2025.
\newblock Grpo-lead: A difficulty-aware reinforcement learning approach for
  concise mathematical reasoning in language models.
\newblock \emph{arXiv preprint arXiv:2504.09696}.

\bibitem[{Zhu et~al.(2025)Zhu, Zhou, Chen, Yu, Ma, Gu, Luo, Tan, and
  Chen}]{zhu2025muq}
Zhu, H.; Zhou, Y.; Chen, H.; Yu, J.; Ma, Z.; Gu, R.; Luo, Y.; Tan, W.; and
  Chen, X. 2025.
\newblock Muq: Self-supervised music representation learning with mel residual
  vector quantization.
\newblock \emph{arXiv preprint arXiv:2501.01108}.

\bibitem[{Ziegler et~al.(2019)Ziegler, Stiennon, Wu, Brown, Radford, Amodei,
  Christiano, and Irving}]{ziegler2019fine}
Ziegler, D.~M.; Stiennon, N.; Wu, J.; Brown, T.~B.; Radford, A.; Amodei, D.;
  Christiano, P.; and Irving, G. 2019.
\newblock Fine-tuning language models from human preferences.
\newblock \emph{arXiv preprint arXiv:1909.08593}.

\end{thebibliography}

\lstset{%
	basicstyle={\footnotesize\ttfamily},% footnotesize acceptable for monospace
	numbers=left,numberstyle=\footnotesize,xleftmargin=2em,% show line numbers, remove this entire line if you don't want the numbers.
	aboveskip=0pt,belowskip=0pt,%
	showstringspaces=false,tabsize=2,breaklines=true}
\floatstyle{ruled}
\newfloat{listing}{tb}{lst}{}
\floatname{listing}{Listing}
%
% Keep the \pdfinfo as shown here. There's no need
% for you to add the /Title and /Author tags.
\pdfinfo{
/TemplateVersion (2026.1)
}
\setcounter{secnumdepth}{2}

\appendix
\section{Demo}
We have provided a Demo about generated songs in the Abstract of the main text. In the Demo, songs generated by the original model and the model optimized by RS, DPO, PPO, and GRPO are presented. The corresponding songs are based on exactly the same prompts, including the lyric prompt and the audio prompt. The generated songs are based on 10 different genres to present model performances in various situations. Furthermore, a video about a hallucination mitigation case is provided. In that video, the location of hallucinations is marked.
\section{Dataset}
\subsection{Preference Dataset}
The preference dataset is composed of music generated by the audio LM. With the help of the language model Hunyuan-large, lyrics are generated. Along with lyric generation, the output is explicitly conditioned on a target musical genre. There are three genres: Sentimental Pop, Rock, Folk, Jazz, Reggae, Contemporary R\&B, Chinese-style Pop, Chinese Traditional Ethnic Music, Chinese-style Opera, and Down-to-earth Songs. From another perspective, audio prompts are 10-second segments with diverse styles. Such audio prompts and generated lyrics are randomly composed to be a complete prompt. The audio prompts are encoded by Codec encoder, and the lyrics are encoded by BPE tokenizer. Finally, the prompts are tokenized and could be processed by the language model. In total, 2857 generated lyrics and 160 audio prompts are prepared for generations. To avoid over-fitting to specific lyrics, each lyric could be combined only with up to 5 random prompts. Then, 21619 prompts are prepared. Since DPO requires paired data sharing the same prompts, each prompt is used to generate 4 songs. Finally, the dataset contains 86746 generated songs. According to the preference data pairing rules described in the main text, these data consist of 25,459 paired chosen-rejected samples, i.e., approximately 1700 hours of generated songs are utilized for off-policy preference optimization.
\subsection{Subjective Evaluation Dataset}
For subjective evaluation, a standardized test set is required. Through systematic screening, we selected 20 lyrics of distinctive songs to comprise the test set. The lyrics are randomly combined with the aforementioned 160 audio prompts to compose prompts. Each under-evaluate model would generate 20 songs based on such prompts. Finally, the generated songs are evaluated by 10 professional music annotators. The evaluation concentrates on 4 dimensions: Overall Quality (OVL), Vocal Melodic Attractiveness (MEL), Vocal-Instrument Harmony (HAM), and Lyrics Following Accuracy (LYC), which will be detailed introduced in Appendix Section C.1. The generated music is asked to be rated from 1 to 5. The higher this score is, the better the performance in the corresponding dimension.
\subsection{Validation Dataset}
Given that the subjective evaluation set contains only 20 data points per model, we constructed an additional 90 prompts disjoint from both the training set and evaluation set to enable more comprehensive model assessment during training. These prompts were generated using the same methodology as employed for the preference dataset. Additionally, the 90 prompts and their corresponding 360 songs generated by the original model (4 songs per prompt) underwent preliminary human annotation. These annotation results served as ground truth for establishing the pairing criteria in the DPO preference dataset, as detailed in Section 3.3 of the main text.

\section{Experimental protocols}
\subsection{Metrics Details}
\subsubsection{PER}
Phoneme Error Rate (PER) is utilized to calculate the reward score in this work. The reward score is formulated as $reward=1-PER$. During RL post-training, the increase in reward score represents a decrease in PER. Moreover, PER is used to evaluate the mitigation of lyric-song hallucinations, i.e., the lyrics alignment of songs. To calculate PER, the vocal track is first extracted by Demucs \cite{defossez2019demucs}, and then Whisperlarge-v2 \cite{radford2023robust} and Zipformer \cite{yao2023zipformer} are utilized for lyric recognition. The utilization of isolated vocal tracks, as opposed to full song audio, significantly enhances recognition accuracy. The Whisper-large-v2 model represents the current state-of-the-art in automatic speech recognition (ASR), demonstrating exceptional performance not only in conventional speech recognition but also exhibiting remarkable robustness in singing voice transcription. Vocal performances commonly contain homophones and elongated vowels that preserve phonemic content. Standard Word Error Rate (WER) measurements consequently register many errors that are imperceptible to human listeners. The phoneme-level PER metric addresses this limitation by more faithfully representing intelligibility and showing higher correlation with subjective lyric clarity ratings.
\subsubsection{Objective Content Score}
As a tool for objective evaluation, Meta Audiobox-Aesthetic \cite{tjandra2025meta} scores the input audios by neural networks. The evaluation metrics are g content enjoyment (CE), content usefulness (CU), production complexity (PC), and production quality (PQ). CE focuses on the subject quality of an audio piece, CU focuses on the likelihood of leveraging the audio as source material for content creation, PC focuses on the complexity of an audio scene, and PQ focuses on the technical aspects of quality instead of subjective quality. In addition, to evaluate the similarity between the generated songs and the audio prompts, a contrastive music–language pre-training model MuQ-MuLan \cite{zhu2025muq} is utilized. The vocal track of the generated songs and the audio prompts are extracted and processed by MuQ-Mulan to calculate an embedding-based similarity.

% Table generated by Excel2LaTeX from sheet 'Sheet3'
\begin{table*}[htbp]
  \centering
  % \caption{Hyper-parameters and training settings for DPO, PPO, and GRPO.}
    \begin{tabular}{cccccl}
    \toprule
    Method & lr    & batch & steps & K & \multicolumn{1}{c}{others} \\
    \midrule
    DPO   & 5e-07 & 3     & 12k   & $NA$    & $\beta$: 0.3 \\
    PPO   & 5e-07 & 16    & 12k   & 4     & $\alpha$: 0.0005, $\gamma$: 1.0, $\lambda$:1.0, $\epsilon$:0.2, entropy: 0 \\
    GRPO  & 5e-07 & 16    & 2k    & 1     & $\beta$: 0, $\epsilon$:0.2, entropy: 0 \\
    \bottomrule
    \end{tabular}%
  \caption{Hyper-parameters and training settings for DPO, PPO, and GRPO.}
  \label{tab1}%
\end{table*}%

\subsubsection{Subjective Musicality Score}
In subjective experiments, the professional musical annotators score the songs from 4 dimensions: Overall Quality (OVL), Vocal Melodic Attractiveness (MEL), Vocal-Instrument Harmony (HAM), and Lyrics Following Accuracy (LYC). Among them, OVL evaluates the overall musicality and naturalness of the generated song, MEL evaluates the beauty, smoothness, and appeal of the vocal melody, HAM evaluates the coherence and integration between vocals and accompaniment, and LYC evaluates the accuracy and clarity with which the vocals follow the intended lyrics.

\begin{table*}[htbp]
  \centering
  % \caption{Subjective results of the baseline model vs. the preference-optimized models. The overall first and second results are marked with \textbf{bold} and \underline{underline}, respectively.}
    \begin{tabular}{l|ccccccc}
    \toprule
    \multicolumn{1}{c|}{\multirow{2}[4]{*}{Methods}} & \multicolumn{7}{c}{\textbf{MOS}$\uparrow$} \\
\cmidrule{2-2}\cmidrule{4-4}\cmidrule{6-6}\cmidrule{8-8}          & \textbf{OVL} &       & \textbf{MEL} &       & \textbf{HAM} &       & \textbf{LYC} \\
    \midrule
    Origin & 3.040$\pm$0.044  &       & 3.455$\pm$0.073  &       & 3.595$\pm$0.068  &       & 2.605$\pm$0.098  \\
    RS    & \underline{3.070$\pm$0.041} &       & \textbf{3.705$\pm$0.063}     &       & \underline{3.615$\pm$0.057}     &       & \underline{2.930$\pm$0.105} \\
    DPO   & 3.020$\pm$0.048  &       & 3.525$\pm$0.071  &       & \textbf{3.650$\pm$0.066}  &       & 2.850$\pm$0.099  \\
    RS+DPO & 3.040$\pm$0.041  &       & 3.560$\pm$0.072  &       & 3.530$\pm$0.069  &       & \textbf{3.160$\pm$0.089}  \\
    PPO   & 2.965$\pm$0.056  &       & 3.330$\pm$0.067  &       & 3.605$\pm$0.069  &       & 2.790$\pm$0.100  \\
    GRPO  & 2.970$\pm$0.041  &       & 3.430$\pm$0.069  &       & 3.460$\pm$0.069  &       & 2.180$\pm$0.053  \\
    \bottomrule
    \end{tabular}%
  \caption{Subjective results of the baseline model vs. the preference-optimized models. The overall first and second results are marked with \textbf{bold} and \underline{underline}, respectively.}
  \label{tab2}%
\end{table*}%

\subsection{Hyper-parameters}
The training settings and hyperparameter settings are listed in Table 1. In Table 1, lr and batch denote the learning rate and the batch size. The number of training steps is also presented. With respect to $K$, it refers to the number of training epochs after each sampling. Since there is no data sampling process in DPO, it's marked as $NA$. While GRPO performs single-epoch optimization per sampled batch, PPO's multi-epoch (K=4) training paradigm inherently increases its total update steps. Unique hyperparameters for each preference optimization are also listed. They are the same as described in the main text. 

\subsection{Training Protocols}
Reinforcement learning methods for LLMs typically require loading multiple data batches or models simultaneously, thus demanding substantial GPU memory. In this work, 8 Nvidia H20 GPUs are utilized for DPO training. The training processes of both PPO and GRPO necessitate the use of 32 Nvidia H20 GPUs to meet their computational demands. Additionally, 8 Nvidia A100 GPUs are used for reward model training. The batch size is set as 3 for DPO training; in other words, 6 pieces of data are processed within a batch since 3 chosen-rejected pairs are included. The batch size for PPO and GRPO is 16 for both. 12k, 12k, and 2k training steps are needed for DPO, PPO, and GRPO, respectively. Finally, the Adam Optimizer is set with $\beta_1=0.9$, $\beta_1=0.95$, and weight decay equals $1e-5$.

\section{Experimental Results}
\subsection{Subjective Experiment Results with 95\% Confidence Intervals}
In subjective experiments, each song is evaluated by 10 professional musical annotators. Thus, multi-rater annotation results require confidence interval-based statistical analysis. In this work, 95\% confidence intervals (CI) are utilized. 95\% CIs are conventionally used to balance precision and reliability in statistical reporting, providing a standard benchmark for comparing results. The subjective experimental results with 95\% CIs are presented in Table 2.

\subsection{Ablation Experiments for PPO}
According to Table 1, there many hyper-parameters to be elaborately tuned for successful PPO training. Thus, PPO with different hyper-parameter setting is evaluated. The validation reward score is shown in Fig. 1, where $\alpha$ and $en$ denote the the weight of KL-divergence loss and entropy loss. $\lambda$ is the accumulating paramter for GAE.

\begin{figure}[htbp]
    \centering
    \includegraphics[width=\linewidth]{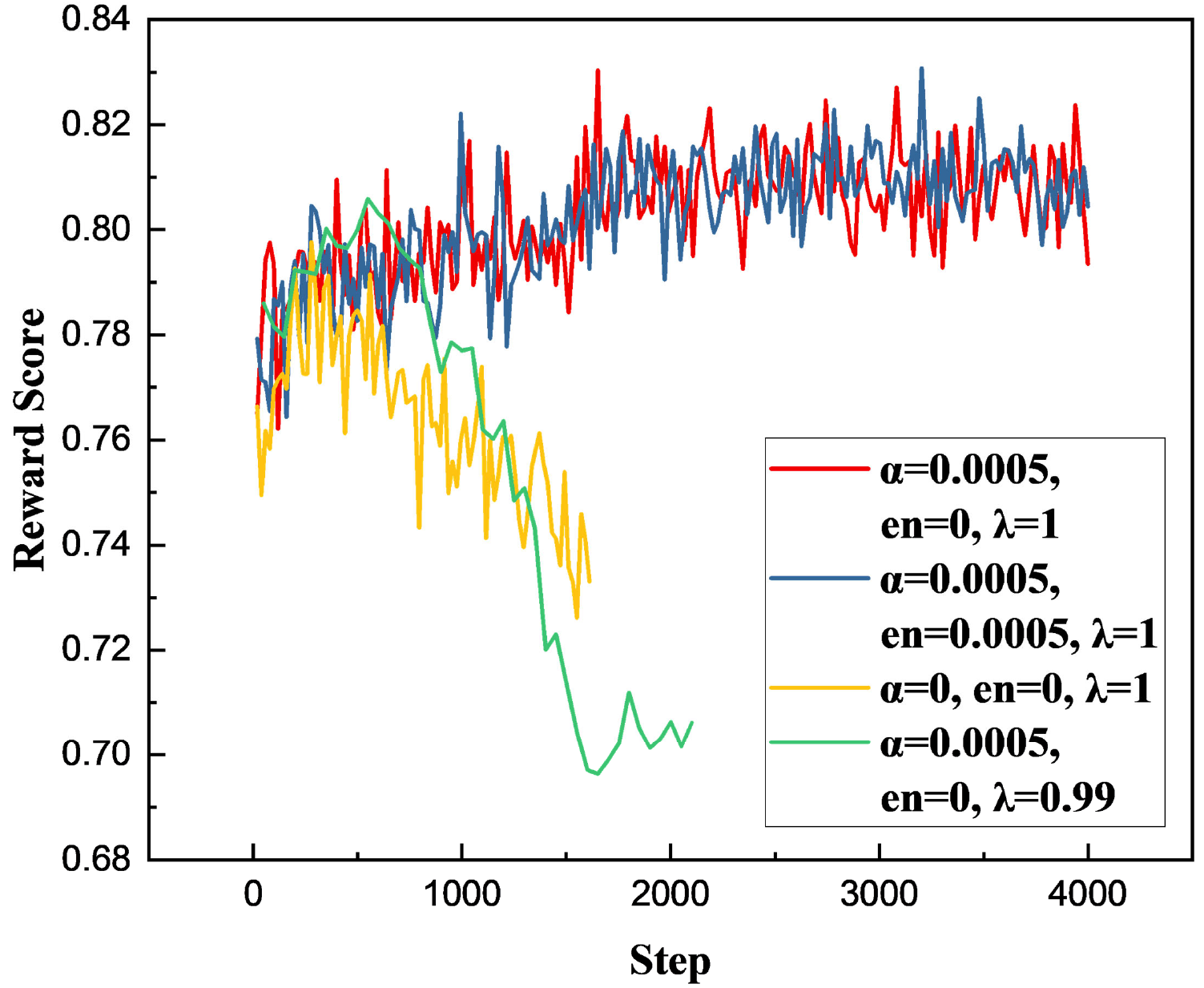}
    \caption{Validation reward score for PPO with different hyper-parameters.}
    % \label{fig:example} % 可选
\end{figure}

According to the results, best practices of PPO are presented. First, $\alpha$ denotes the weight for KL-divergence constraints in PPO training. With such constraints, the PPO training process can prevent the model from deviating excessively from the reference policy, thereby avoiding significant performance degradation. If $\alpha$ is set as 0, the reward score demonstrates a sharp decrease. Secondly, as mentioned in the main text, the $\lambda$ should be set as 1. Otherwise, the terminal tokens used for reward computation would fail to adequately represent the behavioral characteristics of the majority tokens at the sequence onset. As evidenced in Figure 2, even with $\lambda$=0.99, we observe measurable performance degradation. Finally, The entropy bonus in the loss function modulates the balance between exploration and exploitation. It prevents the model from becoming overly conservative and merely generating known high-reward tokens, thereby enhancing its exploratory capabilities. As evidenced by the experimental results in Figure 2, the inclusion of entropy bonus demonstrates negligible impact on model performance. This suggests that for hallucination mitigation tasks, exploration operates within a constrained action space, thereby limiting its overall influence.

\subsection{Ablation Experiments for GRPO}
In the GRPO training framework, we incorporate the token-level loss from DAPO \cite{yu2025dapo}, thereby transforming the reward scoring mechanism from sequence-level to token-level granularity. Since hallucinations manifest at the token level, a single sample may contain multiple hallucinated segments that should contribute equally to the loss function. Experimental results are shown in Fig. 2. For ablation efficiency, this experiment adopts an elevated learning rate (1e-6) compared to the main experiments (5e-7), which accelerates convergence and consequently reduces the total training steps required.

\begin{figure}[htbp]
    \centering
    \includegraphics[width=\linewidth]{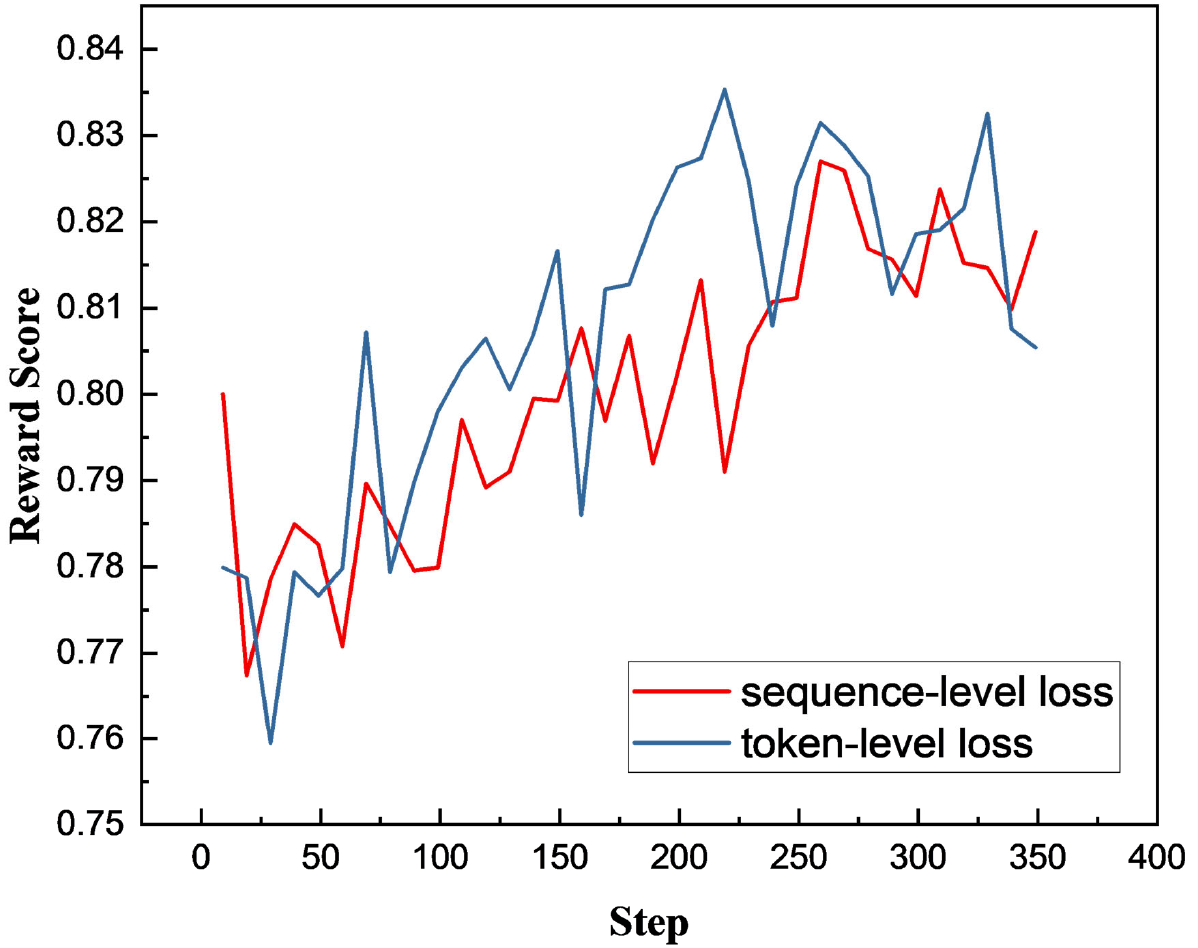}
    \caption{Validation reward score for GRPO with and w/o token-level loss.}
    % \label{fig:example} % 可选
\end{figure}

The experimental results demonstrate that token-level GRPO achieves superior reward scores at equivalent training steps, which aligns with our theoretical analysis regarding the varying significance of different hallucinations.

\subsection{Reward Score Distribution}
In the validation stage, prompts within the validation dataset are utilized to generate music for evaluation. Each model would generate 90 songs whose reward score would be calculated by the reward model. The distribution would reflect the effect of preference optimization, as shown in Fig. 3.  \par 
According to Fig. 3, the observed increase in reward scores post preference optimization indicates effective hallucination mitigation under ideal, unbiased reward modeling conditions. The goal of preference optimization is to address the hallucination phenomenon, specifically to reduce samples with low reward scores (below 0.7). From this perspective, DPO, DPO+RS, and PPO have comparable effects in optimizing the hallucination phenomenon. However, overall, the off-policy DPO and RS+DPO yield a greater improvement in the accuracy of samples, as they increase the generation of more samples with high reward scores (above 0.8).

\begin{figure*}[htbp]
    \centering
    \includegraphics[width=\linewidth]{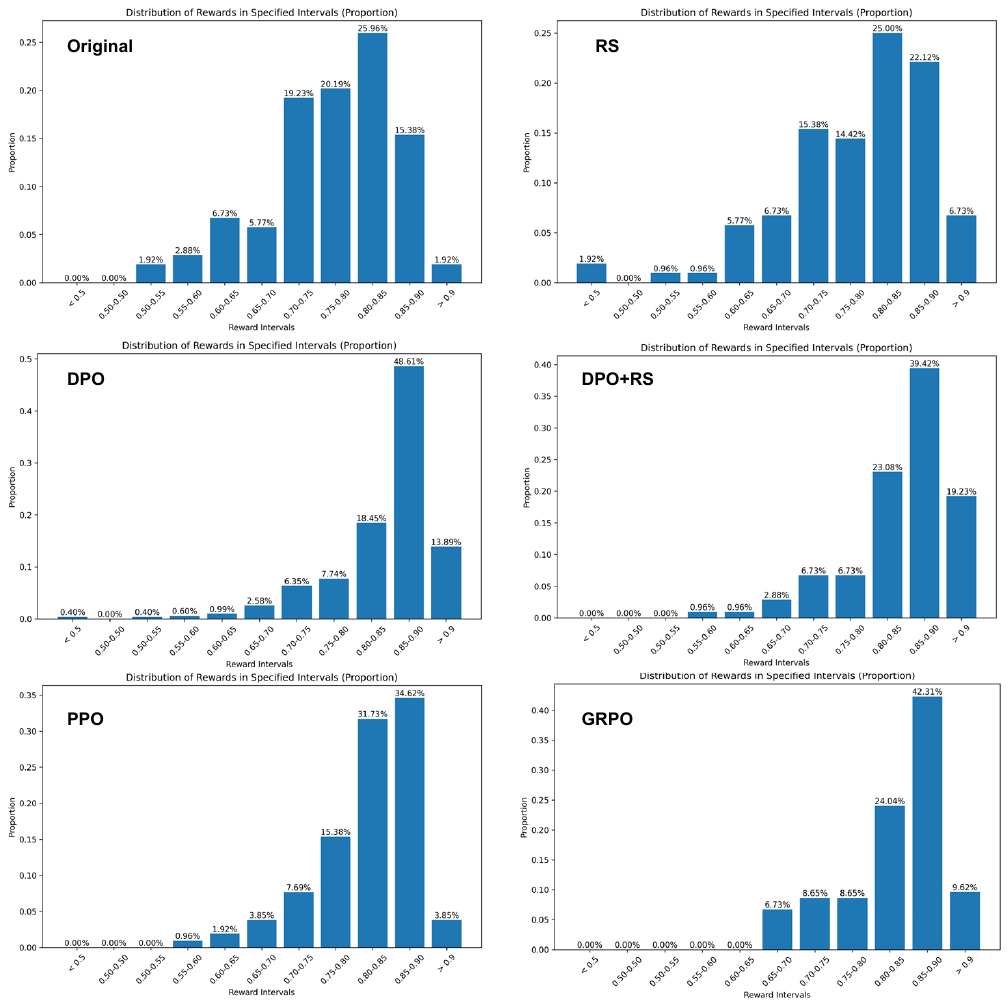}
    \caption{Reward score distribution for original and preference optimized models.}
    % \label{fig:example} % 可选
\end{figure*}
\FloatBarrier

% \bibliography{references}

\end{document}